
\batchmode\font\twelvemsb=msbm10 scaled\magstep1 \errorstopmode
\ifx\twelvemsb\nullfont\def\bb#1{\hbox{\bf#1}}
                        \def\BB#1{\hbox{\bf#1}}
        \message{Blackboard bold not available. Replacing with boldface.}
\else
\font\fourteenmsb=msbm10 scaled\magstep2
\def\bb#1{\hbox{\twelvemsb#1}}
\def\BB#1{\hbox{\fourteenmsb#1}}
\fi

\newfam\scrfam
\batchmode\font\tenscr=rsfs10 scaled \magstep1 \errorstopmode
\ifx\tenscr\nullfont\def\scr{\cal}
        \message{rsfs font not available. Replacing with ``calligraphic''.}
\else
\font\sevenscr=rsfs7 scaled \magstep1
\font\fivescr=rsfs5 scaled \magstep1
\skewchar\tenscr='177 \skewchar\sevenscr='177 \skewchar\fivescr='177
\textfont\scrfam=\tenscr \scriptfont\scrfam=\sevenscr
\scriptscriptfont\scrfam=\fivescr
\def\scr{\fam\scrfam}
\fi

\input phyzzx

\Twelvepoint
\frenchspacing
\hsize=16cm
\vsize=20cm
\voffset=.5cm

\def\noblackbox{\overfullrule=0pt}
\noblackbox

\font\tenrm=cmr10
\font\tenmath=cmmi10

\def\Im{\hbox{Im}\,}
\def\Re{\hbox{Re}\,}
\def\bra{\,<\!}
\def\ket{\!>\,}
\def\II{\hbox{I\hskip-0.6pt I}}

\def\half{{\lower2.5pt\hbox{\tenrm 1}\/\raise2.5pt\hbox{\tenrm 2}}}
\def\ihalf{{\lower2.5pt\hbox{\tenmath i}\/\raise2.5pt\hbox{\tenrm 2}}}
\def\fraction#1{{\lower2.5pt\hbox{\tenrm 1}\/\raise2.5pt\hbox{\tenrm #1}}}
\def\Fraction#1#2{{\lower2.5pt\hbox{\tenrm #1}\/
        \raise2.5pt\hbox{\tenrm #2}}}

\def\tr{\rm tr}

\def\bZ {\bb{Z}}
\def\BZ{\BB{Z}}
\def\bR {\bb{R}}

\def\U {{\scr U}}
\def\H {{\scr H}}
\def\F {{\scr F}}
\def\e {\varepsilon}
\def\a {\alpha}
\def\b {\beta}
\def\d {\delta}
\def\k{\kappa}
\def\l{\lambda}
\def\g {\gamma}
\def\z{\zeta}
\def\L {{\scr L}}
\def\/{\over}
\def\slz{$Sl(2;\bZ)$}
\def\slr{$Sl(2;\bR)$}



\REF\jhs{J.H. Schwarz, {\it Covariant field equations of chiral N=2 D=10
        supergravity}, Nucl. Phys. {\bf B226} (1983) 269.}
\REF\HoweWest{P.S. Howe and P.C. West, {\it the complete N=2, d=10
        supergravity},  Nucl. Phys. {\bf B238} (1984) 181.}
\REF\hulltown{C.M. Hull and P.K. Townsend, {\it Unity of superstring
        dualities}, Nucl. Phys. {\bf B348} (1995) 109.}
\REF\wit{E. Witten, {\it String theory dynamics in various dimensions}, 
Nucl.
        Phys. {\bf B443} (1995) 85.}
\REF\Schwarz{J.H. Schwarz, {\it An $Sl(2;\bZ)$ multiplet of type \II B
        superstrings}, Phys. Lett. {\bf B360} (1995) 13;
        Erratum: ibid. {\bf B364} (1995) 252, hep-th/9508143.}
\REF\Witten{E. Witten, {\it Bound states of strings and p-branes},
        Nucl. Phys. {\bf B460} (1996)  335, hep-th/9510135.}
\REF\pkt{P.K. Townsend, {\it Membrane tension and manifest
        \II B S-duality}, hep-th/9705160.}
\REF\GHMNT{M.T. Grisaru, P.S. Howe, L. Mezincescu, B.E.W. Nilsson and P.K.
        Townsend, {\it N=2 superstrings in a supergravity background},
        Phys. Lett. {\bf 162B} (1985) 116.}
\REF\ceder{M. Cederwall, A. von Gussich, B.E.W. Nilsson and A. Westerberg,
        {\it The Dirichlet super-three-brane in type \II B supergravity},
        Nucl. Phys. {\bf B490} (1997) 163, hep-th/9610148.}
\REF\cedertwo{M. Cederwall, A. von Gussich, B.E.W. Nilsson, P. Sundell
        and A. Westerberg, {\it The Dirichlet super-p-branes in
        type \II A and \II B supergravity}, Nucl. Phys. {\bf B490} (1997) 179,
        hep-th/9611159.}
\REF\bergstown{E. Bergshoeff and P.K. Townsend, {\it Super D-branes}, Nucl.
        Phys. {\bf B490} (1997) 145, hep-th/9611173.}
\REF\agan{M. Aganagic, C. Popescu and J.H. Schwarz, {\it D-brane actions with
        local kappa symmetry}, Phys. Lett. {\bf B393} (1997) 311,
        hep-th/9610249;
        {\it Gauge-invariant and gauge-fixed D-brane actions},
        Nucl. Phys. {\bf B495} (1997) 99.}
\REF\Hull{C.M. Hull, {\it String dynamics at strong coupling},
        Nucl. Phys. {\bf B468} (1996) 113, hep-th/9512181.}
\REF\Vafa{C. Vafa, {\it Evidence for F-theory}, Nucl. Phys. {\bf B469}
        (1996) 403, hep-th/9602022.}
\REF\Bars{I. Bars, Phys. Rev. {\bf D54} (1996) 5203.}
\REF\BLT{P.K. Townsend, {\it Worldsheet electromagnetism and the superstring
        tension}, Phys. Lett. {\bf 277B} (1992) 285;
        E. Bergshoeff, L.A.J. London and P.K. Townsend,
        {\it Space-time scale-invariance and the super-p-brane},
        Class. Quantum Grav. {\bf 9} (1992) 2545, hep-th/9206026.}
\REF\schmidhuber{C. Schmidhuber, {\it D-brane actions}, Nucl. Phys. {\bf B467}
        (1996) 146.}
\REF\Tseytlin{A. Tseytlin, {\it Self-duality of Born--Infeld action and
        Dirichlet 3-brane of type IIB superstring theory},
        Nucl. Phys. {\bf B469} (1996) 51, hep-th/9602064.}
\REF\Green{M.B. Green and M. Gutperle, {\it Comments on three-branes},
        Phys. Lett. {\bf B377} (1996) 28, hep-th/9602077.}
\REF\SchwarzSen{J.H. Schwarz and A. Sen, {\it Duality symmetric actions},
        Nucl. Phys. {\bf B411} (1994) 35, hep-th/9304154.}
\REF\Bengtsson{I. Bengtsson, {\it Manifest duality in Born--Infeld theory},
        hep-th/9612174.}
\REF\Khoudeir{D.S. Berman, {\it SL(2,Z) duality of Born--Infeld theory 
	from self-dual electrodynamics in 6 dimensions}, hep-th/9706208; 
	A. Khoudeir and Y. Parra,
        {\it On duality in the Born--Infeld theory}, hep-th/9708011.}
\REF\Russo{J. Russo, {\it An ansatz for a non-perturbative four-graviton
        amplitude in type IIB superstring theory}, hep-th/9707241.}
\REF\Kallosh{R. Kallosh, {\it Covariant quantization of D-branes},
        hep-th/9705056.}


\Pubnum{ \vbox{ \hbox{G{\"o}teborg-ITP-97-13} \hbox{DAMTP-R/97/38}
\hbox{hep-th/9709002} } }
\date{August, 1997}

\titlepage

\title {\bf The Manifestly $Sl(2;\BZ)$-covariant Superstring}

\author{Martin Cederwall}
\address{Institute for Theoretical Physics\break
G\"oteborg University and Chalmers University of Technology\break
S-412 96 G\"oteborg, Sweden.}
\andauthor{P.K. Townsend}
\address{DAMTP, University of Cambridge,
\break
Silver St., Cambridge CB3 9EW, U.K.}

\abstract{\noindent We present a manifestly \slz-covariant action for the
type \II B superstring, and prove $\kappa$-symmetry for on-shell \II B
supergravity backgrounds.}
\endpage

\chapter{Introduction}
The Type \II B superstring theory has D=10 \II B supergravity as its 
effective
field theory. Until a few years ago, the \slr\ invariance of the
latter [\jhs,\HoweWest] was thought to be an artefact of the field theory
approximation to string theory, but it is now believed that \II B 
superstring
theory is itself an approximation to some underlying non-perturbative 
theory in
which an \slz\ subgroup of \slr\ survives as a symmetry
[\hulltown,\wit]. To the extent to which this theory can be said to be a 
string
theory it describes  an entire \slz\ orbit of `$(p,q)$' strings
[\Schwarz,\Witten] with the (1,0) string being the Green-Schwarz (GS) \II B
superstring and the (0,1) string the D-string. This explains why both the 
usual
\II B superstring action, and that of the D-string, break \slz\
(the action for the `fundamental' string breaks the full group, while
the D-string action, describing $(p,1)$ strings, breaks it to \bZ).
In a recent
paper [\pkt], one of the authors presented a new \slz-covariant string
action that simultaneously describes the entire \slz\ orbit of $(p,q)$
strings. We say `covariant' rather than `invariant' because
an \slz-transformation of the worldsheet fields must be accompanied by an
\slz-transformation of the background. The purpose of this paper
is to present the supersymmetric generalization of this action in a form 
that
makes the \slz\ covariance manifest, i.e. the manifestly
\slz-covariant \II B superstring.

The construction involves establishing the fermionic gauge invariance 
known as
$\kappa$-symmetry. It was shown already in [\GHMNT] that $\kappa$-symmetry 
of
the GS \II B superstring implies constraints on the background that are
equivalent
to the on-shell superspace constraints of \II B supergravity, but this
derivation of them obscures their \slr-invariance. In contrast,
$\kappa$-symmetry of the new \II B superstring action implies the
on-shell \II B
supergravity superspace constraints in \slr-covariant form.
Actually, we shall establish here only that these constraints are 
sufficient
for $\kappa$-symmetry but the results of [\GHMNT] guarantee that they are 
also
necessary. We remark that the \II B on-shell constraints have also been
shown to be
sufficient for $\kappa$-symmetry of the D-3-brane [\ceder], but they have 
not
yet been shown to be necessary (although this is almost certainly the 
case).
Similar results have been found for other \II B
D-branes [\cedertwo,\bergstown,\agan] but the \slr\ invariance of the
background is again obscured. Thus, the action proposed here offers by far 
the
simplest route to a derivation of the \II B supergravity constraints in
manifestly \slr-covariant form as integrability conditions. Another 
motivation
is the potential insight that a manifestly \slz-covariant string
might provide into a conjectural 12-dimensional theory underlying IIB
superstring theory [\Hull,\Vafa,\Bars].

We begin with a review of the string action of [\pkt], rewriting it in a 
form
that makes the \slz\ covariance manifest. The action for the
corresponding manifestly \slz-covariant \II B superstring is
formally identical, but the background is \II B superspace. We prove the
$\kappa$-symmetry of this action subject to the on-shell constraints of 
\II B
supergravity. We conclude with a brief discussion of some potential
applications of our results.

\chapter{The $Sl(2;\BZ)$-invariant String}

The construction of [\pkt] puts together two earlier observations:

({\it i}$\,$)\phantom{\it i} \vtop{\hsize=12.5cm\noindent The tension of a
super $p$-brane may be generated dynamically in a
formulation containing a world-volume $p$-form potential,
which has no local
degrees of freedom [\BLT].}

({\it ii}$\,$) \vtop{\hsize=12.5cm\noindent The Born--Infeld (BI)
field on the world-sheet of the D-string also has no
local degrees of freedom, but the integer quantization of its electric 
field
generates NS-NS charge [\schmidhuber].}

\noindent These two observations make it natural to replace the D-string
tension, equivalent to the magnitude of the RR 2-form charge, by a second 
BI
potential. The two worldsheet gauge fields can be assembled into an $Sl(2)$
doublet, thereby allowing an $Sl(2)$-invariant coupling to the background 
\II B
supergravity fields. The tension of this \slz-invariant string is
generated dynamically; it depends on both the integer RR and NS-NS charges,
both of which arise due to electric field quantization.

Thus, the world-sheet fields comprise not only the target space coordinates
but also an \slr\ doublet $A_r$, $r=1,2$, of abelian gauge
potentials. The latter enter the action via their `modified' 
field-strengths
$$
F_r=dA_r-B_r
\eqn\twoa
$$
where $B_r$ are the pullbacks to the worldvolume of the NS and RR 2-form
potentials. We use the same symbol to denote spacetime forms and their
pullbacks as it should be clear which is intended from the context.
In order to write an \slr-invariant `$F^2$-term', one needs the
background scalars. These belong to the coset $Sl(2;\bR)/SO(2)$ or,
equivalently $SU(1,1)/U(1)$. We shall use the latter description here.
Thus, the scalars are represented by a complex $SU(1,1)$ doublet $\U^r$
($r=1,2$) satisfying the $SU(1,1)$-invariant constraint
$$
\ihalf\e_{rs}\U^r\bar\U^s =1\, .
\eqn\ScalarConstraint
$$
Viewing $\U$ as a $2\times2$ matrix on which $SU(1,1)$ acts from the left,
there is a commuting action of $U(1)$ from the right; we normalize the 
$U(1)$
charge by taking $\U$ to have unit charge. Gauging this $U(1)$ reduces the
number of independent scalars from three to two.

One can move freely between $F_r$ and an $SU(1,1)$-invariant
complex field-strength $\F$ using
$$
\F=\U^r\!F_r\, ,\qquad  F_r=\e_{rs}\Im(\U^s\!\bar\F)\, .
\eqn\twob
$$
Similarly, $H_r=dB_r$ is the \slr\ doublet of NS and RR background 3-form field
strengths, and $\H=\U^rH_r$ the $SU(1,1)$-invariant version. Introducing 
the left-invariant $SU(1,1)$ Maurer--Cartan forms
$$
P=\half\e_{rs}d\U^r\U^s\, ,\qquad  Q =\half\e_{rs}d\U^r\bar\U^s\, ,
\eqn\twod
$$
we can write the Bianchi identity for $\H$ as
$$
D\H+i\bar\H\!\wedge\!P=0\, ,
\eqn\twof
$$
where $D$ is the covariant exterior derivative constructed from the 
$U(1)$ connection
$Q$, which as a consequence of \ScalarConstraint\ is real.

The complex dilaton-axion field of the \II B supergravity background can be
constructed as the projective invariant $\U^2(\U^1)^{-1}=\chi+ie^{-\phi}$. 
It
is sometimes convenient to use the $U(1)$ gauge $\Im\U^1=0$ and identify
$\U^1=e^{\phi\/2}$, $\U^2=e^{\phi\/2}\chi+ie^{-{\phi\/2}}$. Since we want 
to
maintain manifest $Sl(2)$-covariance, we will most of the time  use the 
$\U$'s.

To complete the assembly of ingredients needed for the construction of a
manifestly $SU(1,1)$-invariant action we define the complex scalar density
$$
\Phi=\half\e^{ij}\!\F_{ij}\, .
\eqn\twog
$$
The $Sl(2)$-invariant string action is
$$
S=\half \!\int\!d^2\sigma\,\lambda\bigl(g+\Phi\bar\Phi\bigr)\, .
\eqn\StringAction
$$
The metric is understood to be the pullback of the \slr-invariant
Einstein metric, and $g$ is its determinant, so the action is manifestly
\slr-invariant or, rather, covariant since the invariance of $\Phi$ 
requires
an \slr\ transformation of the background. Note the absence of a
Wess--Zumino term; $\Phi$ already
contains couplings to both the NS-NS and RR sectors.
The action is also invariant under space-time scale transformations:
$$
X\rightarrow\rho X\,,\quad A\rightarrow\rho^2A\,,\quad
        \lambda\rightarrow\rho^{-4}\lambda\,.
$$

We will briefly analyze the action \StringAction\ before moving on to the
supersymmetric case (for a complete demonstration that the equations of
motion implied by
\StringAction\ are those of a $(p,q)$ string we refer to ref. [\pkt]).
The equation of motion for the Lagrange multiplier
$\lambda$ enforces the constraint
$$
g+ \Phi\bar\Phi=0\, .
\eqn\twoi
$$
The equations of motion for the $A_r$ read
$$
d\,\bigl\{\lambda\,\Re(\U^r\bar\Phi)\bigr\}=0\, .
\eqn\twoj
$$
The entities inside the curly brackets must take some constant values.
These are best understood in a canonical framework. The space-components
of the electric fields, the conjugate momenta to $A_r$, are
$$
E^r=\lambda\,\Re(\U^r\bar\Phi)\, .
\eqn\QuantizedE
$$
When the world-sheet is a cylinder, the values of $E^r$ are quantized
to be integers by demanding gauge invariance [\Witten]. We therefore let
$E^r=\e^{rs}n_s$, where $n_r$ is a doublet of integers. This breaks \slr\ 
to
the subgroup \slz. Equation \QuantizedE\ is readily solved by setting
$$
\lambda\Phi=-i\,\U^r\!n_r\, .
\eqn\ComplexTension
$$

In order to identify the tension, we continue the canonical analysis to
the coordinate sector. The momenta are
$$
P_m=\lambda\Bigl(\dot X_m(X')^2-X'_m(\dot X\!\cdot\!X')\Bigr)
        -E^r(B_r)_{mn}X'^n\, ,
\eqn\twok
$$
from which we deduce the following constraints:
$$
\eqalign{
        &0=P\!\cdot\!X'\, ,\cr
        &0=(P_m+E^r(B_r)_{mn}X'^n)^2+(\l\sqrt{-g}\,X')^2\, .\cr}
\eqn\twol
$$
The effective tension is read off from the second of these
equations as
$$
T=\bra\l\sqrt{-g}\ket=\bra|\U^rn_r|\ket\, .
\eqn\twom
$$
It is also clear from \twol\ that $E^r$ are the charges with which the 
string
couples to the background 2-forms $B_r$.
Since the metric entering the action was the \slr-invariant Einstein
metric, this \slz-invariant tension refers to the Einstein frame.
For constant dilaton and axion, and with $n_r=(p,q)$, the tension becomes
$$
T=\sqrt{e^{-\phi}q^2+e^\phi(p+q\chi)^2\,}\,.
\eqn\twon
$$
Rescaling to the string frame by  $(g_{\hbox{\fiverm string}})_{mn}
        =e^{\phi\/2}(g_{\hbox{\fiverm Einstein}})_{mn}$, and
identifying the string coupling constant as $g_s =e^\phi$, the string frame
tension takes the well-known form [\Schwarz]
$$
T_s =\sqrt{\left({q\/g_s}\right)^2+(p+q\chi)^2\,}\, .
\eqn\twoo
$$
Note that the tension and the phase of $\Phi$ naturally
fit together in a `complex tension' $\lambda\Phi$.

\chapter{The $Sl(2;\BZ)$-covariant Superstring}

The action for the manifestly \slz\ superstring is formally identical to
the bosonic action \StringAction, but the `coordinate' worldsheet fields
are now the superspace coordinates 
$Z^M=(X^m,\theta^\mu,\bar\theta^{\bar\mu})$.
Note that the two D=10 chiral spinors of \II B superspace have been
assembled into a single complex chiral spinor. As usual, we introduce the 
frame
1-forms $E^A=(E^a,E^\alpha,E^{\bar\alpha})$ and define the induced 
worldsheet
metric  via the pullback of $E^a$:
$$
g_{ij} = {E_i}^a {E_j}^b\eta_{ab}\, ,
\eqn\threea
$$
where ${E_i}^a=\partial_i Z^M{E_M}^a$.
The 2-form field strengths $F_r$ are as before but the 2-form gauge
potentials $B_r$ are now pulled back from 2-forms on superspace. The
complex 3-form field strength $\H=\U^rdB_r$ is the $SU(1,1)$-invariant superspace field strength introduced in 
[\HoweWest].
We now turn to the issue of $\kappa$-symmetry.

Consider a local fermionic transformation of the type
$$
\d Z^M=\zeta^\a {E_\a}^M+\bar\zeta^{\bar\a} {E_{\bar\a}}^M \, .
\eqn\threeb
$$
The induced variation of the pullback of a superspace form $\Omega$ is 
given by
$$
\d\Omega=\L_\zeta\Omega=(i_\zeta d+di_\zeta)\Omega\, .
\eqn\threec
$$
We use this to calculate that the variation of the $SU(1,1)$-invariant
field-strength $\F$ is
$$
\d\F=-i_\zeta\H-i\bar\F i_\zeta P+i\F i_\zeta Q\,,
                \eqn\FVariation
$$
where we have used $\d A_r =i_\zeta B_r$.
The variation of the induced metric is
$$
\d g_{ij}=2{E_{(i}}^a{E_{j)}}^B\zeta^\a\, {T_{B\a}}^b\eta_{ab} +c.c.
\eqn\threed
$$
where $T_{BC}{}^A$ is the superspace torsion.

At this point we shall use the known on-shell superspace constraints of 
\II B
supergravity. They are [\HoweWest]
$$
\eqalign{
        &\H_{a\a\b}=2i(\g_a)_{\a\b}\, ,\cr
        &\H_{ab\bar\a}=-i(\g_{ab}P)_\a\, ,\cr
        &{T_{\a\bar\b}}^a=i(\g^a)_{\a\b}\, ,\cr
        &{T_{\bar\a\bar\b}}^\g=i{\d_{(\a}}^\g P_{\b)}
                -\ihalf(\g_a)_{\a\b}(\g^aP)^\g\, ,\cr
        &P_{\bar\a}=0\, ,\qquad Q_\a=0\,,\cr}
\eqn\threee
$$
where $P$ in the expressions for the dimension 1/2 parts of $\H$ and $T$
is understood as the spinor component $P_\a$. 
These are the non-vanishing fields at dimensions 0 and 1/2, up to
components that are obtained from these by complex conjugation.
It is straightforward to use these supergravity constraints to obtain
the transformation of the constraint (\twoi): 
At dimension 0 (which is all that is
relevant for bosonic backgrounds) we find
$$
\eqalign{
\left\{\d(g+\Phi\bar\Phi)\right\}_{(0)}
        &=2i\bar E_i\left\{g\g^i\zeta +\Phi\e^{ij}\g_j\bar\zeta\right\}
        +c.c.\cr
&= -2i\bar E_i\g^i \Xi\left\{ \Xi\zeta - \Phi \bar\zeta\right\}
+c.c.}
\eqn\ConstraintTransf
$$
where, as in [\BLT], we have introduced the (real) matrix
$$
\Xi= \half \varepsilon^{ij}\g_{ij}
\eqn\threef
$$
satisfying
$$
\{\Xi,\gamma^i\}=0\qquad  \Xi^2 =-g\, .
\eqn\threeg
$$
At dimension 1/2 (at which we find terms involving background fermions) the
metric has no variation, while
$$
\left\{\d\bar\Phi\right\}_{(1/2)} =-i\bar P_\a(\Xi \zeta
        -\Phi\bar\zeta)^\a\,.
        \eqn\threeh
$$
Thus, the full variation of the constraint, in arbitrary
(on-shell) backgrounds, is
$$
\d(g+\Phi\bar\Phi) = -i(2E_i\g^i\Xi + \Phi \bar P)
(\Xi \zeta - \Phi \bar\zeta) + c.c.
\eqn\threei
$$

If the action is to be invariant the transformation \threei\ must be
cancelled by a variation of the Lagrange multiplier $\lambda$. This means 
that
the expression in \threei\ must vanish modulo the constraint, which
requires $\zeta$ to take the form
$$
\zeta=\Xi\kappa+\Phi\bar\kappa\, ,
\eqn\SolutionForKappa
$$
where the complex chiral spacetime-spinor parameter $\kappa$ is arbitrary.
Given this, we find that
$$
\d(g+\Phi\bar\Phi)= (g+\Phi\bar\Phi)(2iE_i\g^i\Xi\kappa +
i\Phi \bar P\kappa) + c.c.
\eqn\threej
$$
This can clearly be cancelled by a variation of $\lambda$.

To summarize, the action is invariant under the following fermionic gauge
transformations of the worldsheet fields
$$
\eqalign{
\d\lambda &= -i\l(2E_i\g^i \Xi + \Phi \bar P)\kappa + c.c.\cr
\delta Z^M &= (\Xi \kappa + \Phi \bar\kappa)^\alpha E_\alpha{}^M
+(\Xi \bar\kappa + \bar\Phi \kappa)^{\bar\alpha} E_{\bar\alpha}{}^M\cr
\delta (A_r)_i &={E_i}^A (\Xi\kappa+\Phi\bar\kappa)^\a(B_r)_{\a A}
        +c.c. \cr}
\eqn\threek
$$
It can be shown that these $\k$-symmetry transformations effectively
remove half
of the fermionic degrees of freedom, so that there is an on-shell matching 
of
bosonic and fermionic world-sheet fields. This will be demonstrated below 
for
non-vanishing $\Phi$.

\chapter{Discussion}
The action presented in this paper contains in its spectrum the entire
orbit of superstrings coupling with charges $(p,q)$ to the RR and NS-NS
2-form potentials. The spectrum is not confined to coprime pairs of
charges, but a restriction to this irreducible orbit is clearly consistent,
at least at the first-quantized level. Such a restriction preserves the
\slz\ invariance but breaks the scale-invariance. The only sector
preserving scale-invariance is the \slz\ singlet $(p,q)=(0,0)$. This sector
describes a tensionless, or null, superstring, unrelated to the fundamental
superstring by any known duality. It too is removed from
the spectrum by the restriction to the single \slz\ orbit containing the
fundamental string. In this case we may assume that $\Phi$ is non-zero,
and this allows a reformulation of the $\kappa$-symmetry transformations.

In section 3, the complex chiral spinor parameter $\k$ was unconstrained 
but entered into all transformations multiplied by a matrix of half 
maximal 
rank. For non-zero $\Phi$ the constraint imposed by $\lambda$ ensures that 
the induced metric is non-degenerate, so we can divide by $\sqrt{-g}$. 
Setting
$\Phi = e^{i\vartheta}|\Phi|$, we can then rewrite
\SolutionForKappa\ as 
$$
\left[\matrix{\z\cr\bar\z\cr}\right]
=\half({\bf 1}+\Gamma)\left[\matrix{\z\cr\bar\z\cr}\right]\,,
\eqn\foura
$$
where
$$
\Gamma= {1\/\sqrt{-g}}
\left[\matrix{0&e^{i\vartheta}\Xi \cr e^{-i\vartheta}\Xi &0\cr}\right]\, .
\eqn\fourb
$$
Note that $\Gamma^2={\bf 1}$, so that the matrix ${1\/2}({\bf 1}+\Gamma)$ 
is a projector. Since $\tr\Gamma=0$ it projects onto a space of half the 
maximum dimension.

The $(p,q)$ strings treated here represent only part of the BPS spectrum
of the non-perturbative \II B superstring theory. One would like a
manifestly \slz-covariant action for all \II B p-branes. The first case 
to consider after strings is the D-3-brane, which is an S-duality singlet. 
The bosonic action was shown in [\Tseytlin,\Green] to be \slz-covariant in 
the 
sense that an \slz\ transformation of the background can be compensated by 
a world-volume duality transformation of the BI field strength. A {\sl 
manifestly} \slz-covariant formulation of the linearized bosonic action 
has been given in [\SchwarzSen], and elaborated on in 
[\Bengtsson,\Khoudeir]; the essential ingredient is a self-duality 
condition on a complex BI field. The next case to consider is the 
S-duality doublet of solitonic and Dirichlet 5-branes. Given that there 
also exist bound states describing $(p,q)$ 5-branes one would expect there 
to be an \slz-covariant 5-brane action, but there is at present little
indication of how this might be constructed.
   
Finally, it seems possible that one could take the new superstring action 
presented here as the starting point for an \slz-covariant string 
perturbation theory. Note that although all $(p,q)$ strings other than the 
(1,0) 
string are non-perturbative within conventional string theory, the \slz\ 
symmetry is not intrinsically non-perturbative because (in contrast to
the analogous \slz\ symmetry of D=4 N=4 supersymmetric gauge theories) it 
does not exchange electric with magnetic degrees of freedom. Moreover, a 
generalization of the Veneziano amplitude has been proposed in which poles 
correspond to states of $(p,q)$ strings [\Russo]. It would be very 
interesting to investigate whether an \slz-covariant perturbation theory 
based on the action presented here could reproduce this or a similar 
result. The first step towards such a perturbation theory would seem to be 
gauge fixing of the $\k$ symmetry. This can probably be done without 
breaking manifest Lorentz covariance, along the lines of ref. [\Kallosh], 
but it is harder to see how the \slz\ covariance could also be maintained. 
It might be made possible
through the introduction of $SU(1,1)$ harmonics. We leave this to the 
future.

\refout
\end